\documentstyle[amssymb,12pt,epsfig]{article}
\begin{document}
\baselineskip=22pt

\begin{center}
{\Large \bf Effects of the Cosmological Constant as the Origin of
the Cosmic-Ray Paradox}

\bigskip

Zhe Chang\footnote{changz@mail.ihep.ac.cn}~~and~~Shao-Xia Chen\footnote{ruxanna@mail.ihep.ac.cn}\\
{\em Institute of High Energy Physics,
Chinese Academy of Sciences} \\
{\em P.O.Box 918(4), 100039 Beijing, China}\\
Cheng-Bo Guan\footnote{guancb@ustc.edu.cn}\\
{\em ICTS, University of Science and Technology of China}\\
{\em  230026 Hefei, China}
\end{center}

\vspace{1.0cm}
\begin{abstract}
We show that a tiny but non-zero positive cosmological constant,
which has been strongly suggested by the recent astronomical
observations on supernovae and CMBR, may change notably the
behaviors of the ultrahigh energy cosmic ray interacting with soft
photons. The threshold anomaly of the ultrahigh energy cosmic ray
disappears naturally after the effects of the cosmological
constant are taken into account.

\end{abstract}
%
\vspace{1.0cm}
 PACS numbers: 98.70.Sa, ~95.85.Pw, ~03.30.+p, ~98.80.Es.
\vspace{1.0cm}

Recently there is great interest in the study of the ultrahigh
energy cosmic ray (UHECR) and the TeV-photon paradoxes. Hundreds
of events with energies above $10^{19}$eV and about 20 events
above $10^{20}$eV have been observed \cite{data1}-\cite{data5}.
All these observed events exceed the Greisen-Zatsepin-Kuzmin (GZK)
\cite{GZK} threshold. In principle, the photopion production
process by the cosmic microwave background radiation (CMBR) should
decrease the energies of these protons to the level which is below
the corresponding threshold. The second paradox \cite{tev-review}
comes from the detected 20TeV photons from the Mrk 501 (a BL Lac
object at a distance of 150Mpc). Similar to the case of UHECR, due
to interaction with the IR background photons, the 20TeV photons
should have disappeared before arriving at the ground-based
detectors. The two puzzles have a common feature that both of them
can be considered to be some threshold anomalies: energy of an
expected threshold is reached but the threshold is not observed.
There are yet numerous published suggestions \cite{UHECR}-\cite{
solution1} for possible solution of the UHECR and the TeV-$\gamma$
paradoxes. In particular, authors \cite{loren1}-\cite{loren8} have
suggested that Planck scale physics and the violation of the
Lorentz invariance can be the origin of these anomalies. Along
this way, noncommutative geometry \cite{a2} and doubly special
relativity approaches \cite{DSR1}-\cite{DSR4} for the cosmic-ray
paradox have been set up and many interesting results have been
obtained. However, all of these investigations are far beyond the
standard cosmology theory and the standard model of particle
physics.

The most important progress made in cosmology in recent years is
that the astronomical observations on supernovae
\cite{constant-super-1, constant-super-2} and CMBR
\cite{costant-cmbr} show that about two third of the whole energy
in the universe is contributed by a small positive cosmological
constant. An asymptotic de Sitter (dS) spacetime is promised
naturally. The physics in an asymptotic dS spacetime has been
discussed extensively \cite{ds-paper-1}-\cite{ds-paper-3}.

In this Letter, starting from solving equations of motion for a
free particle in the dS sapcetime, we try to give a reasonable
solution to the cosmic ray paradox by taking the effects of the
cosmological constant into account.

The dS spacetime can be described as a submanifold of a five
dimensional pseudo-Euclidean space
\begin{equation}
\label{xi}
(\xi^0)^2-(\xi^1)^2-(\xi^2)^2-(\xi^3)^2-(\xi^5)^2=-\frac{1}{\lambda}~,
\end{equation}
\begin{equation}
\label{metric}
ds^2=(d\xi^0)^2-(d\xi^1)^2-(d\xi^2)^2-(d\xi^3)^2-(d\xi^5)^2~,
\end{equation}
where $\lambda$ is the curvature of the dS spacetime. This
realization of dS spacetime is invariant obviously under the
action of the dS group $SO(4,~1)$.

It is convenient to study the kinematics in dS spacetime by
introducing the Beltrami coordinates $x^i$,
\begin{equation}
\label{xi-x}
x^i\equiv\frac{\xi^i}{\sqrt{\lambda}\xi^5}~,~~~~~(i=0,~1,~2,~3)~.
\end{equation}
In the Beltrami coordinate system, the dS spacetime can be
rewritten into the form
\begin{equation}
\label{sigma}
\sigma\equiv\sigma(x,~x)=1-\lambda\eta_{ij}x^ix^j>0~,
\end{equation}
\begin{equation}
\label{sigma-metric}
ds^2=\left(\frac{\eta_{ij}}{\sigma}+\frac{\lambda\eta_{ir}\eta_{js}x^rx^s}{\sigma^2}\right)dx^idx^j~,
\end{equation}
where $\eta_{ij}={\rm diag}(1,~-1,~-1,~-1)$. It is easy to check
that Eqs. (\ref{sigma}) and (\ref{sigma-metric}) are invariant
under transformations of the $SO(4,~1)$
\begin{equation}
\begin{array}{l} \label{x-x-tilde}
x^i\rightarrow\tilde{x}^i=\sigma(a,~a)^{1\over2}\sigma(a,~x)^{-1}(x^j-a^j)D^i_j~,\\
[0.5cm]
D^i_j=L^i_j+\lambda\left(\sigma(a,~a)+\sigma(a,~a)^{1\over2}\right)^{-1}\eta_{kl}a^la^iL^k_j~,\\
[0.5cm] L\equiv(L^i_j)~~\in SO(3,~1)~,\\
[0.5cm]\sigma(a,~a)>0~.
\end{array}
\end{equation}
We notice that there is a subgroup $SO(4)$ of the de Sitter one
$SO(4,1)$, which consists of spatial transformations among
$x^\alpha$ ($\alpha=1,~2,~3$). It is not difficult to show that
$\xi^0$
 $(\equiv\sigma(x,x)^{-1/2}x^0)$ is invariant under the spatial
transformations. Thus, we can say that two spacelike events are
simultaneous if they satisfy

\begin{equation}
\label{samultaneous} \sigma(x,~x)^{-\frac{1}{2}}x^0=\xi^0={\rm
constant}~.
\end{equation}
Therefore, it is convenient to discuss physics of the dS spacetime
in the coordinate $(\xi^0,~x^\alpha)$. In this coordinate, the
metric can be rewritten into the form

\begin{equation}
\label{diag-metric}
ds^2=\frac{d\xi^0d\xi^0}{1+\lambda\xi^0\xi^0}-(1+\lambda\xi^0\xi^0)
\left[\frac{d\rho^2}{(1+\lambda\rho^2)^2}
+\frac{\rho^2}{1+\lambda\rho^2}d\Omega^2\right]~,
\end{equation}
where $\rho^2\equiv\displaystyle\Sigma{x^\alpha x^\alpha}$ and
$d\Omega^2$ denotes the metric on the sphere $S^2$. \\
If a proper time $\tau$ is introduced as

\begin{equation}
\tau\equiv
\frac{1}{\sqrt{\lambda}}\sinh^{-1}(\sqrt{\lambda}\xi^0)~,
\end{equation}
one can get a Robertson-Walker-like metric

\begin{equation}
ds^2=d\tau^2 -\cosh^2(\sqrt{\lambda}\tau)
\left[\frac{d\rho^2}{(1+\lambda\rho^2)^2}
+\frac{\rho^2}{1+\lambda\rho^2}d\Omega^2\right]~.
\end{equation}
The Casimir operator of the de Sitter group can be used to express
the one-particle states

\begin{equation}
\label{invariant}
\left(\frac{1}{\sqrt{-g}}\partial_i\left(\sqrt{-g}g^{ij}\partial_j\right)
+m^2_0\right)\Phi(\xi^0,x^\alpha)=0~,
\end{equation}
where $\Phi(\xi^0,x^\alpha)$ denotes a scalar field or a component
of vector field for a particle with given spin $s$.

Making use of the diagonal metric $(\ref{diag-metric})$, we can
rewrite the de Sitter invariant operator in the following form

\begin{eqnarray}
\frac{1}{\sqrt{-g}}\partial_i\left(\sqrt{-g}g^{ij}\partial_j\right)&=&\left(
1+\lambda\xi^0\xi^0
\right)\partial_{\xi^0}^2+4\lambda\xi^0\partial_{\xi^0}   \\
&&-\left( 1+\lambda\xi^0\xi^0 \right)^{-1}\left[
\left(1+\lambda\rho^2 \right)^2
\partial^2_\rho+2\rho^{-1}\left( 1+\lambda\rho^2 \right)^2
\partial_\rho \right] \nonumber\\
&&+\left( 1+\lambda\xi^0\xi^0 \right)^{-1}\left( 1+\lambda\rho^2
\right)\rho^{-2} \left[-\partial^2_{\bf u}+s(s+1)\right]~,
\nonumber
\end{eqnarray}
where $\partial^2_{\bf u}$ denotes the Laplace operator on $S^2$.

To solve the equation of motion, we write the field
$\Phi(\xi^0,x^\alpha)$ into the form
\[
\Phi(\xi^0,\rho,{\bf u})=T(\xi^0)U(\rho)Y_{lm}({\bf u})~.
\]
This form of the field transforms the equation of motion
into\cite{ds-paper-3}

\begin{eqnarray}
&&
\left[(1+\lambda\xi^0\xi^0)^2\partial^2_{\xi^0}+4\lambda\xi^0(1+\lambda\xi^0\xi^0)
\partial_{\xi^0}+m_0^2(1+\lambda\xi^0\xi^0)+(\varepsilon^2-m_0^2)
\right]T(\xi^0)=0~,
\nonumber \\[0.5cm]
&& \left[\partial^2_\rho+
\displaystyle\frac{2}{\rho}\partial_\rho-\left[\frac{
m_0^2-\varepsilon^2}
{(1+\lambda\rho^2)^2}+\frac{l(l+1)+s(s+1)}{\rho^2(1+\lambda\rho^2)}\right]
\right]U(\rho)=0~, \nonumber\\ [0.5cm]
&&\left[\partial^2_{\bf u}+l(l+1)\right]Y_{lm}({{\bf u}})=0,
\end{eqnarray}
where $Y_{lm}({{\bf u}})$ is the spherical harmonic function and
$\varepsilon$ is a constant.

Solutions of timelike part of the field are of the forms

\begin{equation}
T(\xi^0)\sim\left(1+\lambda\xi^0\xi^0\right)^{-1/2} \cdot\left\{
\begin{array}{l}
P_{\nu}^{\mu}(i\sqrt{\lambda}\xi^0)~, \\
\\
Q_{\nu}^{\mu}(i\sqrt{\lambda}\xi^0) ~,%
\end{array}
\right.
\end{equation}
where $\mu,~\nu~$ satisfy

$$
\begin{array}{l}
\nu(\nu+1)=2-\lambda^{-1}m_0^2~, \\[0.5cm]
\mu^2=1+\lambda^{-1}(\varepsilon^2-m_0^2) ~.%
\end{array}%
$$
For the radial equation of the field, we can write the solutions
as the form

\begin{equation}
U(\rho)\sim\rho^l(1+\lambda\rho^2)^{k/2}F\left(\frac{1}{2}(l+s+k+1),~\frac{1}{2}
(l+s+k),~l+s+\frac{3}{2};~-\lambda\rho^2\right)~,
\end{equation}
where $k$ denotes the radial quantum number
$$k^2-2k-\lambda^{-1}(\varepsilon^2-m_0^2)=0~.$$
To be normalizable, the hypergeometric function in the radial part
of the field has to break off, leading to the quantum condition

\begin{equation}
\frac{l+s+k}{2}=-n~,~~~~~(n\in \mathbf{N})~.
\end{equation}
Then, we get the dispersion relation for a free particle in the dS
spacetime

\begin{equation}\label{ppp}
E^2=m_0^2+\varepsilon'^2+\lambda(2n+l+s)(2n+l+s+2)~.
\end{equation}
The dispersion relation (\ref{ppp}) combined with the conservation
laws forms a powerful and elegant means of treating the kinematics
in the collision and decay processes in the spacetime with a
positive cosmological constant.

We first consider the head-on collision between a soft photon of
energy $\epsilon$, momentum ${\bf q}$ and a high energy particle
of energy $E_1$, momentum ${\bf p}_1$, which leads to the
production of two particles with energies $E_2$, $E_3$ and momenta
${\bf p}_2$, ${\bf p}_3$, respectively. From the laws of
conservation of energy and momentum, we have
\begin{equation}
\label{threshold1} E_1+\epsilon=E_2+E_3~,
\end{equation}
\begin{equation}
\label{threshold2} p_1-q=p_2+p_3~.
\end{equation}
In the C. M. frame, $m_2$ and $m_3$ are at rest at threshold, so
that they have the same velocity in the lab frame. It's easy to
give the following relation
\begin{equation}
\label{p3p2} \frac{p_2}{p_3}=\frac{m_2}{m_3}.
\end{equation}

For the process of the UHECR interacting with the CMBR photons,
$$p+\gamma\rightarrow p+\pi~,$$
we obtain the threshold
\begin{equation}
\label{UHECR-threshold}\displaystyle E^{\rm UHECR}_{{\rm
~th},~\lambda}\simeq\frac{(m_{N}+m_{\pi})^2-m_{N}^2
+\lambda^{*}\left(1+\frac{m_N}{m_{\pi}}+\frac{m_{\pi}}{m_N}\right)}
{2\left(\epsilon+\sqrt{\epsilon^2-\lambda^{*}}\right)}~,
\end{equation}
where $\lambda^{*}=\lambda(2n+l+s)(2n+l+s+2)~$, and we have used
the dispersion relation (\ref{ppp}) and popular approximated
relations for relativistic particles
\begin{equation}
\label{12}
\epsilon^2=q^2+\lambda^{*}=q^2+\lambda(2n+l+s)(2n+l+s+2)~,
\end{equation}
\begin{equation}
\label{13} E_i=\sqrt{m^2_i+p^2_i+\lambda_i^{*}}\simeq
p_i+\frac{m^2_i}{2p_i}+\frac{\lambda^{*}_i}{2p_i}~,~~(i=1,~2,~3)~.
\end{equation}
It should be noticed that the $\lambda$ dependent term in the
expression of the threshold (\ref{UHECR-threshold}) can not be
omitted in spite of the tiny value of the observed cosmological
constant $\lambda (\simeq 10^{-85}{\rm GeV}^2)~$. The reason is
that the angular momentum, which appears in the dispersion
relation of a free particle in dS spacetime, is a cosmological
quantity. The distance between the particle and the coordinate
origin is at the level of $\frac{1}{\sqrt{\lambda}}$.

In {\bf FIG.1}, we give a plot of the dependence of the threshold
$E^{\rm UHECR}_{{\rm ~th},~\lambda}$ on the value of the curvature
($\lambda$) of the dS spacetime.

%
%
\begin{center}

\includegraphics[height=80mm,width=90mm]{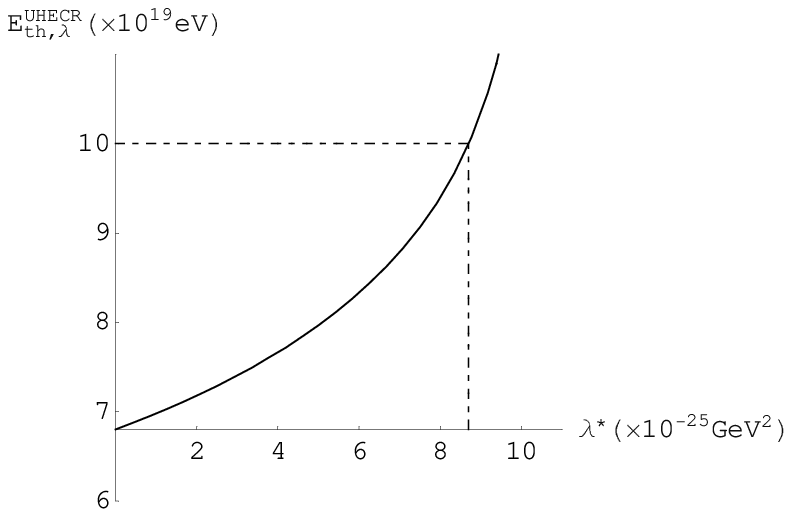}

{\bf FIG.1}~~~~ {\scriptsize The cosmological constant
($3\lambda$) dependence of the threshold $E^{\rm UHECR}_{{\rm
~th},~\lambda}$ in the interaction process between
\hspace*{1.9cm}the UHECR protons and the CMBR photons
$(10^{-3}{\rm eV})~$. \hspace*{\fill}}
\end{center}

From {\bf FIG.1}, we know clearly that a tiny but non-zero
positive cosmological constant increases the threshold sharply in
the photopion production process of the CMBR photons with the
ultrahigh energy cosmic ray. For the observed cosmological
constant, if the CMBR takes the quantum number $(2n+l)$ to be
about $10^{30}$, the energies of all the observed UHECR events are
below the theoretical threshold and the threshold anomaly
disappears.

In the interaction process between the TeV-$\gamma$ ray and the IR
background photons,
$$\gamma+\gamma\rightarrow e^++e^-~,$$we have similar dispersion
relations as
\begin{equation}
\label{22}
\epsilon^2=q^2+\lambda^{*}=q^2+\lambda(2n+l+s)(2n+l+s+2)~,
\end{equation}
\begin{equation}
\label{23} E_i=\sqrt{m^2_i+p^2_i+\lambda_i^{*}}\simeq
p_i+\frac{m^2_i}{2p_i}+\frac{\lambda^{*}_i}{2p_i}~,~~(m_1=0)~.
\end{equation}
The threshold we obtained is of the form
\begin{equation}
\label{Tev-threshold} E^{\gamma}_{{\rm
~th},~\lambda}\simeq\frac{2m_e^2+3\lambda^{*}}{\epsilon+\sqrt{\epsilon^2-\lambda^{*}}}~.
\end{equation}
We present a plot of the curvature ($\lambda$) dependence of the
threshold $E^{\gamma}_{{\rm ~th},~\lambda}$ in {\bf FIG.2}.
%
%
\begin{center}
\includegraphics[height=80mm,width=90mm]{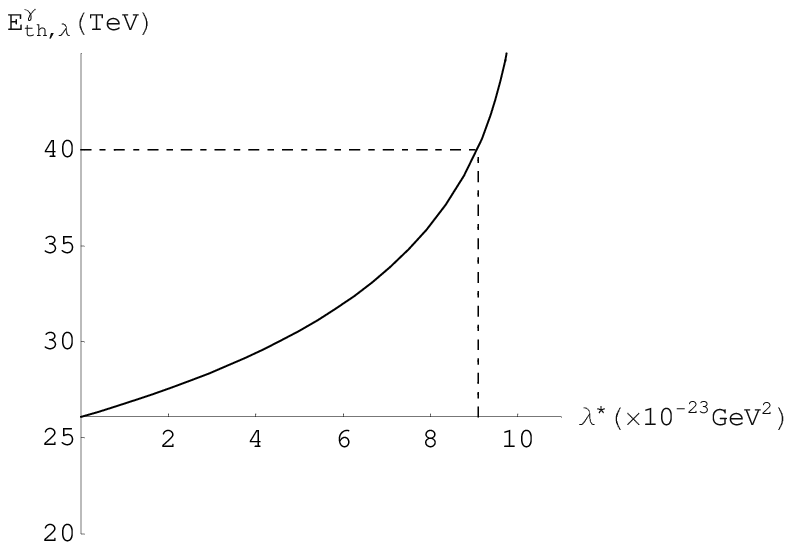}

{\bf FIG.2}~~~~ {\scriptsize The cosmological constant
($3\lambda$) dependence of the threshold $E^{\gamma}_{{\rm
~th},~\lambda}$ in the process of the TeV-$\gamma$
\hspace*{2.45cm}interacting with IR background photons
$(10^{-2}{\rm eV})~$.\hspace*{\fill}}
\end{center}

From {\bf FIG.2}, one can see that the threshold $E^{\gamma}_{{\rm
~th},~\lambda}$ is also very sensitive to the varying of the
cosmological constant if the IR background also has a quantum
number $(2n+l)$ about $10^{31}$. Similar to UHECR, we now have a
new theoretical threshold of $40{\rm TeV}$ for this process and
the threshold anomaly doesn't confuse us any more.

In this Letter, we have investigated kinematics in the de Sitter
spacetime and obtained a deformed dispersion relation for free
particles. In particular,  the CMBR and IR background interacting
with extremely high energy cosmic rays have been presented in the
framework. We noticed that the familiar GZK cutoff might be
deviated from if the effects of the cosmological constant were
taken into account.  Therefore, the cosmological constant may be
the origin of the threshold anomaly of the cosmic ray. Of course,
the origin of the UHECR is one of the outstanding puzzles of
modern astrophysics\cite{origin1}--\cite{origin3}. Today's
understanding of the phenomena responsible for the production of
UHECR is still limited. Furthermore, it is well-known that to
solve the so-called flatness problem, the horizon problem and
magnetic monopoles in the standard big-bang theory, we should
introduce the inflation models\cite{inflation}. In fact, recent
observations of WMAP show strong evidence for
inflation\cite{wmap}. After all of these factors have been dealt
with carefully, a more reliable scenario of the threshold anomaly
of the cosmic ray can be obtained.

\vspace*{0.5cm}
%
\centerline{\large\bf Acknowledgements} The work was supported in
part by the Natural Science Foundation of China. One of us
(C.B.G.) is supported by grants through the ICTS (USTC) from the
Chinese Academy of Sciences.

%

\end{document}